\documentstyle[12pt,amsfonts]{article}
\newcommand{\mbf}[1]{\mbox{\boldmath$ #1$}}
\newcommand{\be}{\begin{equation}}
\newcommand{\ee}{\end{equation}}
\newcommand{\ba}{\begin{eqnarray}}
\newcommand{\ea}{\end{eqnarray}}

\begin{document}

\begin{center}
{\LARGE\bf The role of electromagnetic trapped modes in extraordinary transmission
in nanostructured materials}

\vskip 0.7cm
{\large A.G. Borisov ${}^{a,b}$, F.J. Garc\'{i}a de Abajo ${}^{b,c}$, and \ \ S.V.
Shabanov ${}^{d,b}$}

\vskip 0.7cm

${}^{a}$\ {\it Laboratoire des Collisions Atomiques et Mol\'{e}culaires, UMR\\
CNRS-Universit\'{e} Paris-Sud 8625, B\^{a}t. 351, Universit\'{e} Paris-Sud,\\
91405 Orsay CEDEX, France}\\
$^{b}$\ {\it Donostia International Physics Center (DIPC), Apartado Postal 1072,\\
20080 San Sebasti\'{a}n, Spain\\
$^{c}$\ Centro Mixto CSIC-UPV/EHU, Apartado Postal 1072, 20080 San\\
Sebasti\'{a}n, Spain}\\
${}^{d}$\ {\it Department of Mathematics, University of Florida, Gainesville, FL\\
23611, USA}

\end{center}

\begin{abstract}
We assert that the physics underlying the extraordinary light transmission
(reflection) in nanostructured materials can be understood from rather
general principles based on the formal scattering theory developed in
quantum mechanics. The Maxwell equations in passive (dispersive and
absorptive) linear media are written in the form of the Schr\"{o}dinger
equation to which the quantum mechanical resonant scattering theory (the
Lippmann-Schwinger formalism) is applied. It is demonstrated that the
existence of long-lived quasistationary eigenstates of the effective
Hamiltonian for the Maxwell theory naturally explains the extraordinary
transmission properties observed in various nanostructured materials. Such
states correspond to quasistationary electromagnetic modes trapped in the
scattering structure. Our general approach is also illustrated with an
example of the zero-order transmission of the TE-polarized light through a
metal-dielectric grating structure. Here a direct on-the-grid solution of
the time-dependent Maxwell equations demonstrates the significance of
resonances (or trapped modes) for extraordinary light transmission.
\end{abstract}

\begin{description}
\item  PACS: 42.79.Dj, 41.20.Jb, 42.25.Fx, 03.65.Nk


\end{description}

\newpage

\section{Introduction}

Supported by technological progress, studies of the interaction of
electromagnetic radiation with nanostructured materials have become an area
of intense research driven by potential applications in optics and photonics 
\cite{1}--\cite{4}. In particular, it has been found that metal \cite
{3}--\cite{16} and dielectric \cite
{17}-\cite{24}\ grating structures posses extraordinary
transmission (reflection) properties within narrow intervals of wavelengths
close to the grating period. While for dielectric gratings a common point of
view on this phenomenon, as occurring due to the existence of trapped modes
or guided wave resonances \cite{18}-\cite{24}, seems to be
established, there is still an ongoing discussion on the origin of a nearly
100\% light transmission within narrow wavelength range(s) observed in slit
and hole arrays made in metal films. Although similar results are obtained
with different theoretical approaches, in which the Maxwell equations are
numerically solved, an explanation of the underlying physics, as due to
excitations of coupled surface plasmons and/or cavity modes \cite{5,6},
competes with the dynamical diffraction theory point of view \cite{9,10}.
Remarkably, all available studies of metal grating structures have been
carried out with the TM-polarized light (the magnetic field is parallel to
the slits) where surface plasmons can indeed be excited.

Transmission and reflection properties of grating structures are typically
studied by stationary methods in the frequency domain. Nevertheless, the
dynamics of light scattering on gratings can partly be guessed from
stationary results. Indeed, consider a femtosecond (broad band) pulse
impinging on a grating structure such that the spectral range of the pulse
is much larger than the grating transmission window(s). From the uncertainty
principle it follows that, in order for transmission of light to occur
only within a narrow frequency range, the radiation should last much longer
than the duration of the initial pulse. This implies that the corresponding
electromagnetic modes have to be trapped by the nanostructured material
during a sufficiently long time. Such modes are known in scattering
theory as scattering resonances \cite{25}.

Here we propose a general point of view on the physics underlying the
extraordinary light transmission in nanostructured materials. By
reformulating the Maxwell equations in the form of the Schr\"{o}dinger
equation and by using quantum scattering theory, we show that this
phenomenon is a direct consequence of the existence of trapped
electromagnetic modes possessing large lifetimes. Based on this general
principle, various structures can be custom-designed that would transmit or
reflect light within a designated narrow wavelength range(s). As an example
to illustrate our approach, transmission properties of a metal-dielectric
grating structure are studied for  TE-polarized light (the electric field
is parallel to the slits) impinging normally on the grating. A direct
on-the-grid solution of the time-dependent Maxwell equations demonstrates
the significance of resonances (or trapped modes) for the enhanced light
transmission. Interestingly enough, such a system has never been studied
before, probably because of the absence of the coupling of electromagnetic
modes to plasmons, often thought to be the main mechanism of the
extraordinary light transmission. It should be understood that our choice of
the geometry does not imply an attempt to favor either the plasmon or
cavity--mode point of view in, generally, strongly coupled systems.

\section{Metal--Dielectric Grating}

We begin with an example of a grating structure sketched in Fig. 1 to
illustrate our basic idea. The grating structure has translational symmetry
along one of the Euclidean axes, chosen to be the $y$ axis. The structure is
periodic along the $x$ axis with period $D_{g}$, and the $z$ direction is
transverse to the structure. For the sake of comparison with previous works 
\cite{5}, the parameters are chosen to be: the grating period $%
D_{g}=1.75\,\mu m$, the thickness (along the $z$ axis) $h=0.8\div 1.4\,\mu m$%
, and the opening width $a=0.3\div 0.35\,\mu m$. In the case of the TM
radiation impinging on the grating with no dielectric fillings, the
extraordinary transmission properties have already been reported and
analyzed \cite{5,7,10,11,13}. The question arises whether enhanced
transmission can be obtained for the TE radiation, or according to our
remarks in the previous section, whether or not trapped modes exist and can
be excited.

To obtain a rough estimate of the wavelengths of possible trapped (or
quasi-stationary) electromagnetic modes in the system, consider first the
case of a perfect metal grating. For a moment, we also neglect the effects
due to a finite thickness of the grated metal slab in the $z$-direction. In
other words, we neglect the coupling between trapped modes and radiation
modes outside the grating. Then the $x$ component of the wave vector is
quantized as a consequence of the zero boundary conditions at the
metal-dielectric interface. A quantization of the $z$ component of the wave
vector can be understood as Fabry-Perrot modes in a dielectric slab inside
the metallic waveguide, i.e., the modes for which the dielectric slab is $%
100\%$ transparent. Admissible (quantized) values of the wave vector inside
the grating define wavelengths $\lambda _{nm}$ of trapped electromagnetic
modes that can be excited by the incident radiation 
\begin{equation}
\varepsilon \left( \frac{2}{\lambda _{nm}}\right) ^{2}=\left( \frac{n}{a}%
\right) ^{2}+\left( \frac{m}{h}\right) ^{2}\ ,  \label{1}
\end{equation}
where $\varepsilon $ is the dielectric constant, and $n,m=1,2,...$. Note
that, since $a<h$ for the grating geometry, the first term in the r.h.s. of
Eq. (\ref{1}) is the most relevant for the threshold to excite quasistationary
states (standing waves). In what follows we are interested in the zero
diffraction mode for wavelengths $\lambda \geq D_{g}$ so that the reflected
and transmitted beams propagate along the $z$-axis. In this case the
threshold for trapped modes to exist is determined by the lowest mode, $%
n=m=1 $, leading to the condition 
\begin{equation}
\varepsilon \geq \frac{1}{4}\left( \left( \frac{D_{g}}{a}\right) ^{2}+\left( 
\frac{D_{g}}{h}\right) ^{2}\right) \ .  \label{2}
\end{equation}
Taking into account the grating parameters, condition (\ref{2}) can be
satisfied only if $\varepsilon \geq 9.7$ for the range of $a$ and $h$
specified above. Thus, if there is no dielectric filling ($\varepsilon =1$),
the grating structure would totally reflect the TE radiation with $\lambda
\geq D_{g}$, which is indeed supported by our numerical simulations. This is
in contrast to the TM radiation case where $100\%$ transmission can be
reached for the very same grating with no dielectric fillings and made of
the perfect conductor \cite{5}.

In our numerical study, we consider the grating with openings filled with a
non-dispersive dielectric with $\varepsilon =11.9$ (this corresponds to Si
within the wavelength range under consideration). The metal is described by
the Drude model:\ 
\begin{equation}
\varepsilon _{M}(\omega )=1-\frac{\omega _{p}^{2}}{\omega ^{2}+i\omega
\gamma }\ .  \label{3}
\end{equation}
For the sake of comparison with previous works the plasma frequency and the
attenuation are taken as: $\omega _{p}=9eV$, and $\gamma =0.1eV$ \cite{5}.
Transmission and reflection properties of the grating structure are
calculated by means of the wave packet propagation method. The method is
based on the representation of the Maxwell equations in the form of the
Schr\"{o}dinger equation for which the initial value problem is numerically
solved by a time-stepping algorithm. A detailed description can be found
elsewhere \cite{23,24}.

For the specific case considered here, the Hamiltonian formalism is as
follows. Let ${\bf D}={\bf E}+{\bf P}$ where ${\bf D}$, ${\bf E}$, and ${\bf %
P}$ are the electric induction, the electric field, and the medium
polarization vector, respectively. In the Drude model the medium
polarization vector satisfies the second order differential equation 
\begin{equation}
\ddot{{\bf P}}+\gamma \dot{{\bf P}}=\omega _{p}^{2}{\bf E}\ ,  \label{4}
\end{equation}
where derivatives with respect to time $t$ are denoted by overdots. Equation 
\ref{4} must be solved with zero initial conditions, ${\bf P}=\dot{{\bf P}}%
=0 $ at $t=0$. Define an auxiliary field ${\bf Q}$ by $\dot{{\bf P}}=\omega
_{p}{\bf Q}$. The Maxwell's equations are cast in the Schr\"{o}dinger form: 
\begin{equation}
i\dot{\Psi}=H\Psi \ ,  \label{5}
\end{equation}
where the wave function $\Psi $ and the Hamiltonian $H$ are 
\begin{equation}
\Psi =\pmatrix{ \sqrt{\varepsilon }{\bf E} \cr {\bf B} \cr {\bf Q}}\ ,\ \ \
\ \ H=\pmatrix{ 0 & ic\varepsilon ^{-1/2}\mbf{\nabla }\times & -i\omega _{p}
\cr -ic\mbf{\nabla }\times \ \varepsilon ^{-1/2} & 0 & 0 \cr i\omega _{p} &
0 & -i\gamma}\ .  \label{6}
\end{equation}
The norm of the wave function, $\Vert \Psi \Vert ^{2}=\int d{\bf r}\Psi
^{\dagger }\Psi $, is proportional to the total electromagnetic energy of
the wave packet \cite{23,26}. When the attenuation is not present, $\gamma
=0 $, the Hamiltonian is Hermitian, and the norm (energy) is conserved. The
quantities $\varepsilon $ and $\omega _{p}$ are position dependent so that $%
\varepsilon =1$ everywhere outside the dielectric inclusions, and $\omega
_{p}=0$ everywhere outside the metal part of the grating.

It follows from Eq. (\ref{5}) that $\Psi (t+\Delta t)=\exp (-i\Delta tH)\Psi
(t)$. In our simulations, the action of the infinitesimal evolution operator 
$\exp (-i\Delta tH)$ on the wave function $\Psi $ is carried out by the
algorithm described in \cite{23}. The initial wave packet is Gaussian and
propagates along the $z$ axis perpendicular to the grating. Its spectrum is
broad enough to cover the frequency range of interest. A change of variables
is used to enhance the sampling efficiency in the vicinity of medium
interfaces so that the boundary conditions are accurately reproduced by the
Fourier-grid pseudospectral method \cite{27,28}. A typical size of the mesh
corresponds to $-15D_{g}\leq z\leq 15D_{g}$, and $-0.5D_{g}\leq x\leq
0.5D_{g}$ with $512$ and $128$ knots, respectively. The frequency resolved
transmission and reflection coefficients are obtained via the
time-to-frequency Fourier transform of the signal at some distance in front
and behind the grating \cite{29}. An absorbing layer is introduced at the
grid boundaries in order to suppress artificial reflections of the wave
packet \cite{CAPS}.

\section{Results and discussion}

In Fig. 2a we show an interpolated image of the time evolution of the
electric field $E_{y}$ along the $z$-axis passing through the center of the
grating (see Fig. 1). The grating structure is characterized by $\varepsilon
=11.9$, $D_{g}=1.75\,\mu m$, $a=0.35\mu m$ and $h=1.4\mu m$. The red and
blue colors correspond to positive and negative values of the field,
respectively. The horizontal axis represents the $z$-coordinate expressed in
units of the grating period, $D_{g}$. The grating extends from $z=0$ to $%
z=h/D_{g}$. The vertical axis represents the propagation time measured in
femto-seconds. The initial pulse impinging on the grating has a duration of
approximately $25fs$. The instant when the pulse hits the grating followed
by the main reflected signal is clearly visible in the figure. One also
observes that a fraction of the electromagnetic energy is stored in the
grating structure and leads to a long lasting radiation on both the
transmission and reflection sides. This lasing effect extends to a 
pico-seconds time range, i.e., it is much longer than the duration of the initial pulse. It
can be explained as due to the existence of trapped electromagnetic modes or
resonances. It is the radiation of decaying trapped modes that comes with
a phase opposite to the corresponding harmonic in the initially reflected
pulse to the left from the grating structure and leads finally to the
reduced reflection. The same lasing effect to the right from the grating
structure is responsible for high transmission at the same frequency (see
further discussion and results in Fig. 3).

Figure 2b shows the dynamics of the electric field $E_y$ in the same
setting, but the attenuation $\gamma$ is set to zero. In this case, the
trapped modes do not dissipate their energy into the metal. As a result,
they live longer, which is clearly seen from comparison of the color
intensity of the vertical strip in the middle of Figs. 2a and 2b (that
represents the electric field of the trapped modes).

The calculated transmission coefficient is presented in Fig. 3 as a function
of the wavelength expressed in units of the grating period, $D_{g}$. Each of
the resonances observed in Fig. 3 is associated with the corresponding
trapped mode. The rough estimate given in Eq.(\ref{1}) of their energies
(frequencies) can be improved by taking into account the penetration of the
field into the metal whose dielectric properties are described by Eq. (3).
This yields 
\begin{equation}
\omega _{m}=\sqrt{\Omega ^{2}+\left( \frac{\pi cm}{h\sqrt{\varepsilon }}%
\right) ^{2}}\ ,\ \ \ m=1,2,...\ ,  \label{7}
\end{equation}
where $\Omega $ is the frequency of the lowest eigenmode in the stationary
equation 
\begin{equation}
\frac{\partial ^{2}E(x)}{\partial x^{2}}+\frac{\Omega ^{2}}{c^{2}}%
\varepsilon _{M}(x,\Omega )E(x)=0\ .  \label{8}
\end{equation}
Equation (\ref{8}) is solved numerically under the condition that $E(x)$ must
decay exponentially outside the interval $0<x<a$ (i.e., in the metal). For $%
a=0.35\mu m$ and $h=1.4\mu m$, this gives the following resonant wavelengths
(expressed in the units of grating period): $\lambda _{1}=1.493$, $\lambda
_{2}=1.352$, $\lambda _{3}=1.185$, $\lambda _{4}=1.031$, $\lambda _{5}=0.899$%
. The improved estimate of the resonant wavelengths agrees closely with the
results obtained from numerical simulations for the exact problem only for
the largest resonant wavelength ($\lambda _{1}=1.493$). The wavelengths
corresponding to the maximum of the transmission coefficient for higher
modes are redshifted as compared to the estimated values. A similar result
was reported by Takakura \cite{7}, but for the TM incident wave
polarization. The red-shift can be explained by spreading of the trapped
modes into the vacuum due to a finite thickness of the grating (see also
Fig. 4 and its discussion below), while the Fabry-Perrot modes of the
electromagnetic field, used in our rough estimate, satisfy the zero boundary
condition at the dielectric-vacuum interface. Clearly, an increase of the
spatial volume occupied by a standing wave implies increasing its wavelength
and, hence, lowering its frequency.

Because of dissipative losses of energy in the Drude metal, the
transmittance does not reach 100\% and is, in fact relatively small. While
for a lossless medium the sum of the reflection and transmission
coefficients must be one as follows from the electromagnetic energy
conservation, this is not the case when attenuation is present (blue curve
in Fig. 3). The maximal loss of energy corresponds to resonant wavelengths.
This can be easily understood because the trapped modes remain in contact
with the metal much longer than the main pulse (cf. Figs. 2a and 2b), and,
therefore, can dissipate more energy through exciting surface electrical
currents in the metal. We further illustrate this point by computing the
transmission coefficient in the same system but without attenuation ($\gamma
=0$). The result is shown by the dashed black curve, which reaches $1$ at
resonant wavelengths. Observe the deviation of the transmission coefficient
from 1 for the narrowest resonance at $\lambda \sim 1.5D_{g}$ even in the
absence of absorption. This resonance possesses an extremely long lifetime,
so much so that we had to stop the calculation before it had decayed
completely; 
that is, the total energy trapped into this mode was not completely radiated out
and, hence, was not fully accounted for. For the resonance at $\lambda
<D_{g} $ the transmission coefficient does not reach 1 because we study only
the zero diffraction order scattering channel.

Wave functions (field configurations) of the trapped modes can be extracted
from the time-dependent wave packet by the time-to-frequency Fourier
transform: 
\begin{equation}
\Psi (\omega )=\int\limits_{0}^{\infty }\Psi (t)\;e^{i\omega t}\;dt\ .
\label{9}
\end{equation}
The wave function of a particular trapped mode is obtained by setting $%
\omega $\ to the frequency at which the transmission coefficient attains the
corresponding maximum. In order to improve the contrast, the wave functions
have been extracted for the case with no attenuation. The results are
presented in Fig. 4 for the grating with parameters $\varepsilon =11.9$, $%
D_{g}=1.75\,\mu m$, $a=0.35\mu m$, and $h=1.4\mu m$. There are five colored
strips in Fig. 4. Each colored strip represents the electric field of the
trapped mode associated with corresponding maxima of the transmission
coefficient (see Fig. 3). The data (from the top to the bottom) starts with
the largest wavelength resonance at $\lambda \approx 1.5D_{g}$ and ends
with the resonance at $\lambda \approx 1D_{g}$. The red and blue colors
represent, respectively, negative and positive values of the electric field
amplitude. Each colored strip covers the coordinate range: $-0.26\mu m\leq
x\leq 0.26\mu m$ (along the vertical axis) and the range for $z$ (horizontal
axis) is specified in the figure in units of $D_{g}$. The fields localized
inside the grating and fields radiated into the vacuum (the lasing effect)
are clearly visible in the figure. The trapped modes localized inside the
dielectric part of the grating exhibit a nearly Fabry-Perrot pattern with
respect to the quantization in the $z$-direction. Observe a slight spreading
of the field into the vacuum regions, $z<0$ and $z>h$, which explains the
redshift of the resonant wavelengths as compared to the pure Fabry-Perrot
prediction give by Eq. (\ref{7}).

The structure of the field in the present case is such that one can regard
the grating openings as an ensemble of independent emitters. They are
coherently excited by the incident pulse, and their coherent emission builds
up the radiation field associated with the resonantly enhanced transmission
(reflection) properties of the grating. This is in contrast with the
previously reported TM results, where the excitation of plasmons leads to
the coupling between effective emitters associated with the grating
openings. This point is further illustrated in Fig. 5, where we show the
transmission coefficient calculated for the grating structure with
parameters $\varepsilon =11.9$, $a=0.3\mu m$, $h=0.8\mu m$, and different
periods $D_{g}=D_{0}\equiv 1.75\mu m$, $D_{g}=D_{0}/1.5$, and $D_{g}=D_{0}/4$%
. The change of the grating period does not affect positions of the peaks in
the transmission coefficient, pointing at the independence of the trapped
field associated with different openings. On the contrary,
for the TM polarized light, the resonant wavelengths are
``pinned'' to the grating period \cite{5,6,11} -- this fact being a
reason for the ongoing discussion on the role of surface plasmons in the
TM radiation transmission. The overall increase of the transmission
coefficient when the grating period is reduced is due to the increase of the
density of emitters (openings), while the time scale of the lasing effect
remains the same because it is set by the attenuation of the metal and by
the coupling of each individual grating region to the vacuum. In agreement
with Eq. (\ref{1}), a direct comparison of the result for $D_{g}=D_{0}$ with
the results presented in Fig. 3 shows that the reduction of$\ h$, and
primarily of $a$, leads to the blue shift of the whole resonance series.
Similar dependence of the resonance wavelengths on the grating thickness $h$
has been reported as well for the TM polarization \cite{5,6}. However, some
caution is needed when comparing the TE and TM results in view of 
their different boundary conditions.

\section{The significance of trapped modes in resonant scattering}

The example of the grating structure considered above suggests that the
knowledge of long-lived trapped modes is crucial for a custom design of
nanostructured materials with enhanced transmission (reflection) properties
in designated narrow intervals of wavelengths. Here we offer a rather
general approach which establishes a direct relation between transmission
(reflection) properties of nanostructured materials and the existence of
trapped modes. We assert that once the Maxwell equations have been
reformulated in the Schr\"{o}dinger form, the significance of trapped modes
for the light transmission can immediately be understood from the basic
principles of quantum resonant scattering theory \cite{35}.

Recall that the approach relies on a representation of the total wave
function of the system as a sum of non-resonant and resonant contributions 
\cite{30,31}. Consider the case without losses so that the corresponding
Hamiltonian is Hermitian. Let $H$ be the total Hamiltonian of a
nanostructured material, supporting resonance(s), and $H_{0}$ be a
Hamiltonian responsible for a non-resonant scattering. In the present case $H
$ and $H_{0}$ are Hamiltonians of the metal grating with and without
dielectric insertions, respectively. For the TM polarization the natural
choice will be to set $H_{0}$ to be a Hamiltonian of the simple metal slab
with no gratings. The Lippmann-Schwinger formalism \cite{32,33} is applied
to describe the scattering of a plane wave on a scatterer that has resonant
excitations. Now we will show that the existence of long-lived
quasi-stationary states in the symmetric ($z\longrightarrow -z$
transformation) grating with dielectric insertions implies that there exists
a frequency at which the grating becomes transparent. If $\omega $ is the
frequency of the incoming wave, then a solution of the Schr\"{o}dinger
equation $(H-\omega )\Psi =0$ can be written in the form 
\begin{equation}
\Psi =\Psi _{0}+G^{+}(\omega )\;\left( H-H_{0}\right) \Psi _{0}\equiv \Psi
_{0}+\Psi ^{+}\ ,  \label{10}
\end{equation}
where $G^{+}(\omega )=[\omega -H+i\eta ]^{-1}$, $\eta \rightarrow 0+$ is the
Green function, $\Psi _{0}$ is a solution of $(H_{0}-\omega )\Psi _{0}=0$,
and $\Psi ^{+}$ satisfies radiation (outgoing wave) boundary conditions and
describes the scattered wave due to dielectric insertions. If the incoming
wave is polarized along the grating, then the wave functions $\Psi _{0}$ and 
$\Psi ^{+}$ contain only one component of the electric field which is
denoted $E_{0}$ and $E^{+}$, respectively. Let the frequency $\omega $ be
in the range in which the structure described by $H_{0}$ Hamiltonian is a
total reflector, then 
\begin{equation}
E_{0}\rightarrow A_{0}e^{ikz}+{A}_{0}^{\ast }e^{-ikz}\ ,\ \ \ \ z\rightarrow
-\infty   \label{11}
\end{equation}
and $E_{0}$ vanishes as $z\rightarrow \infty $ assuming that the grating is
centered at $z=0$. For the TM polarization, the magnetic field should be
considered along similar lines.

Quasi-stationary states, or resonances, that exist in the grating with
dielectric insertions correspond to eigenvectors of $H$ with outgoing wave
boundary conditions and, therefore, they are associated with poles of the
Green function $G^{+}(\omega )$. Note that due to the complex boundary
conditions $H$ would have complex eigenvalues $\omega _{0}-i\Gamma /2$ with
negative imaginary parts ($\Gamma >0$) which specify resonance widths. Thus,
in the vicinity of a pole, the frequency dependence of the Green function
can be approximated by 
\begin{equation}
G^{+}(\omega )\sim \frac{1}{\omega -\omega _{0}+i\Gamma /2}\ .  \label{12}
\end{equation}
Let $\omega $ be near a resonant frequency (within the resonance width),
while the resonant frequency $\omega _{0}$ is assumed to remain in the range
of total reflectivity of the pure metallic grating. In this case, from the
symmetry of the Hamiltonian $H$ under the parity transformation, $%
z\rightarrow -z$, it follows that 
\begin{equation}
E^{+}\rightarrow \lbrack {\rm sign}(z)]^{p}A^{+}\,e^{\pm ikz}\ ,\ \ \ \
z\rightarrow \pm \infty \ ,  \label{13}
\end{equation}
where the parity factor of $A^{+}$ corresponds to either symmetric $(p=0)$
or anti-symmetric $(p=1)$ eigenfunction of $H$. The lowest frequency
resonance corresponds to the symmetric solution. The amplitude $A^{+}$ has
to be found from the energy flux conservation. The incident flux is $%
|A_{0}|^{2}$. The outgoing flux is $|A_{0}\pm A^{+}|^{2}+|A^{+}|^{2}$. Let $%
\phi _{0}$ and $\phi ^{+}$ be the phases of $A_{0}$ and $A^{+}$, respectively.
Then from the flux conservation we infer that $|A^{+}|^{2}=|A_{0}|^{2}\cos
^{2}(\phi _{0}-\phi ^{+})$. From Eq. (\ref{12}) it follows that $\phi ^{+}$,
as a function of the frequency $\omega $, rapidly changes by $\pi $ over a
small interval containing the resonant frequency (the eigenvalue of $H$),
while the phase $\phi _{0}$ describing the non-resonant scattering is nearly
constant, or changes slowly. Therefore by continuity of $\phi ^{+}$ in the
vicinity of the resonant frequency there exists a frequency at which $%
|A_{0}|^{2}=|A^{+}|^{2}$, that is, the incoming flux coincides with the
transmitted flux and the grating becomes transparent.

Thus, in the absence of attenuation, the existence of trapped
electromagnetic mode(s) or resonances necessarily leads to  100\%
transmission close to the wavelengths of the trapped modes in symmetric
nanostructured materials which otherwise (in the absence of such modes) are
not transparent. The same approach can be used to analyze a possible 100\%
reflection in dielectric gratings, which otherwise (in the absence of
trapped modes) are nearly transparent in the zero diffraction order. It also
explains possible Fano profiles in the transmission (reflection) coefficient
in the cases when non-resonant scattering described by the Hamiltonian
$H_0$ leads to both reflection and transmission. 

In a generic case, in order for 100\% transmission to be possible,
the parity symmetry of the Hamiltonian (system) is not required, but
the weaker condition (\ref{13}) on the asymptotic behavior of $\Psi^+$
is indeed necessary. The latter readily follows from the flux
conservation. A sufficient condition for 100\% transmission is the 
absence of scattering channels with different quantum numbers  
at the resonance frequency, e.g., higher order diffraction and/or
coupling to scattering states with different polarization.
All these effects lead to breaking of the flux conservation in
a selected scattering channel. If the coupling to other channels
is not significant, an enhanced (not 100\%) transmission can 
still be observed in the selected channel.  
It can be quantified by conventional
means of quantum scattering theory applied to the effective
electromagnetic Hamiltonian of the system in question.
In practice, gratings often have the imperfections which induce 
a coupling between TE and TM polarization. Thus, in the TE resonant
scattering channel  100\% transmission will be lost due to leaking
of the energy flux into higher order diffraction  and/or TM
polarization channels induced by the imperfections. 
One should
not, however, expect a resonant transmission in the TM channel
facilitated by the imperfections since the resonant (enhanced) 
transmission is essentially due to the constructive interference
which, in turn, occurs thanks to the periodicity of the structure,
while imperfections are usually randomly distributed. In fact,
in a lossless grating, a deviation of the transmission coefficient
from one at the resonant frequency can be used as a measure of the
grating quality.

Finally, it is worth mentioning that a
loss of the electromagnetic energy in dispersive materials
prevents  100\% transmission (reflection) to occur, again
because of  breaking of the flux
conservation even in a single scattering channel available.
This is clearly seen from our numerical results with and
without  attenuation (Fig. 3). Our formalism offers 
a possibility to quantify such effects by studying the 
unitarity violation of the scattering matrix caused by
the skew-Hermitian part of the effective Hamiltonian
(see, e.g., \cite{arno}). Note that in quantum systems a leak of the
probability density into scattering channels weakly coupled to the one
of interest is often modelled by an effective non-Hermitian Hamiltonian
for the main (selected) scattering channel(s) only.

\section{Conclusions}

We have elucidated the role of trapped modes in the extraordinary light
transmission in nanostructured materials by reformulating the Maxwell
equations for passive linear media in the form of the time dependent
Schr\"{o}dinger equation and applying to the latter the basic principles of
 quantum resonant scattering theory, in particular, the
Lippmann-Schwinger formalism. Trapped electromagnetic modes in
nanostructured materials play the same role as resonances in quantum
scattering. This offers well developed quantum mechanical techniques to
study resonant light transmission and reflection properties of gratings and
other nanostructured materials. We have illustrated this approach by a
detailed numerical study of a metal-dielectric grating. In particular, for
the TE polarization of the normal incident radiation, the grating, while
being a total reflector in the zero diffraction order when no dielectric
fillings are present, has been shown to become transparent for certain
(resonant) wavelengths when the fillings are present. In accord with the
quantum resonant scattering theory, stationary states have been observed in
the latter case and none in the former.

\vskip 1cm {\bf Acknowledgments}

A.G.B. and S.V.S. thank the DIPC for the support and hospitality.
Useful discussions with Dr. G. G\'omez-Santos are gratefully acknowledged.
S.V.S. is grateful to Prof. J.R. Klauder for fruitful and stimulating
discussions, and Drs. R. Albanese and T. Olson for their continued support.

\vskip 1cm


\newpage

{\bf Figure captions}

Fig. 1.\qquad Schematic representation of the studied system. The radiation
is incident along the normal to the slab (the $z$-axis), consisting of a
grating structure with alternating regions of metal and dielectric along the 
$x$-direction. The dark grey and shaded regions correspond, respectively,
to the metal and dielectric parts of the structure. The system is
translationally invariant along the $y$-direction.\bigskip

Fig. 2a.\qquad (Color online) Interpolated image of the time evolution of
the electric field $E_{y}$ along the $z$-axis passing through the center of
a dielectric region of the grating (see Fig. 1). The grating structure is
characterized by $\varepsilon =11.9$, $D_{g}=1.75\,\mu m$, $a=0.35\mu m$,
and $h=1.4\mu m$. The red and blue colors correspond, respectively, to
positive and negative values of the field with the color intensity related
to the field magnitude. The horizontal axis represents
the $z$-coordinate expressed in units of the grating period, $D_{g}$. The
grating extends from $z=0$ to $z=h/D_{g}$. The vertical axis represents the
propagation time measured in femto-seconds. The metal is described by the
dielectric function of Eq. 3.\bigskip

Fig. 2b.\qquad (Color online) The same as Fig. 2a, but with no damping inside
the metal ($\gamma =0$).\bigskip

Fig. 3.\qquad (Color online) Zero-order transmission coefficient as a
function of the wavelength of the incident radiation measured in units of
the period $D_{g}$. The calculation is carried out for the grating structure
characterized by $\varepsilon =11.9$, $D_{g}=1.75\,\mu m$, $a=0.35\mu m$,
and $h=1.4\mu m$. The solid red and dashed black curves correspond to
calculations done with ($\gamma =0.1eV$) and without ($\gamma =0$) damping
in the metal, respectively. The sum of the reflection and transmission
coefficients for $\gamma =0.1eV$ is shown as the blue curve. Its deviation
from $1$ represents the loss of electromagnetic energy because of the
absorption in the metal.\bigskip

Fig. 4.\qquad (Color online) Electric field of the trapped modes inside the
grating as function of $x$ and $z$ coordinates. The results are presented
for the grating with parameters $\varepsilon =11.9$, $D_{g}=1.75\,\mu m$, $%
a=0.35\mu m$, and $h=1.4\mu m$. Each colored strip represents the electric
field of the trapped mode associated with corresponding maxima of the
transmission coefficient (see Fig. 3). The data (from top to bottom)
starts with the largest wavelength resonance at $\lambda \approx
1.5D_{g}$ and ends with the resonance at $\lambda \approx 1D_{g}$. The
red and blue colors correspond, respectively, to positive and negative
values of the field. The data has been normalized to 1 at maximum so that
color scale covers the range $[-1,+1]$. The $x$-range for each strip (along
the vertical axis) corresponds to $-0.26\mu m\leq x\leq 0.26\mu m$ and
the range
for $z$ (horizontal axis) is specified in the figure in units of $D_{g}$.
\bigskip

Fig. 5.\qquad (Color online) Zero-order transmission coefficient as a
function of the wavelength of the incident radiation measured in units of
the period $D_{0}\equiv 1.75\mu m$. The computed data are given for a
grating structure characterized by $\varepsilon =11.9$, $a=0.3\mu m$, $%
h=0.8\mu m$, and different periods $D_{g}$. The red curve shows the
transmission coefficient for $D_{g}=D_{0}\equiv 1.75\mu m$, the blue curve
for $D_{g}=D_{0}/1.5$, and the black curve for $D_{g}=D_{0}/4$.


\end{document}